\documentclass[%
  reprint,
  amsmath,
  amssymb,
  aps,
  prb,
  floatfix,
  showkeys
]{revtex4-2}

\usepackage{graphicx}
\usepackage{dcolumn}
\usepackage{bm}
\usepackage[dvipsnames]{xcolor}
\usepackage[colorlinks=true, allcolors=gray]{hyperref}
\usepackage{float}
\usepackage{verbatim}
\usepackage{xspace}
\usepackage{siunitx}
\usepackage[version=4]{mhchem}

\graphicspath{{./}{./figures}}

\newcommand{\contract}[0]{\odot^c}
\newcommand{\contractc}[1]{\odot^{#1}}
\newcommand{\model}[0]{CAMP\xspace}
\newcommand{\sifull}[0]{Supplementary Information\xspace}
\newcommand{\sishort}[0]{SI\xspace}

\newcommand*{\fref}[1]{Fig.~\ref{#1}}
\newcommand*{\tref}[1]{Table~\ref{#1}}
\newcommand*{\eref}[1]{Eq.~\eqref{#1}}

\newcommand*{\olcite}[1]{Ref.~\cite{#1}}
\newcommand*{\method}[1]{\hyperref[#1]{Methods}}
\newcommand*{\discussion}[1]{\hyperref[#1]{Discussion}}

\begin{document}

\title{Cartesian Atomic Moment Machine Learning Interatomic Potentials}

\author{Mingjian Wen$^1$}
\email{mjwen@uestc.edu.cn}
\author{Wei-Fan Huang$^2$}
\author{Jin Dai$^2$}
\author{Santosh Adhikari$^2$}
\affiliation{$^1$Institute of Fundamental and Frontier Sciences, University of Electronic Science and Technology of China, Chengdu, 611731, China \\
$^2$Department of Chemical and Biomolecular Engineering, University of Houston, Houston, TX, 77204, USA
}

\begin{abstract}

  Machine learning interatomic potentials (MLIPs) have substantially advanced atomistic simulations in materials science and chemistry by balancing accuracy and computational efficiency.
  While leading MLIPs rely on representing atomic environments using spherical tensors, Cartesian representations offer potential advantages in simplicity and efficiency.
  Here, we introduce the Cartesian Atomic Moment Potential (CAMP), an approach to building MLIPs entirely in Cartesian space.
  CAMP constructs atomic moment tensors from neighboring atoms and employs tensor products to incorporate higher body-order interactions, providing a complete description of local atomic environments.
  Integrated into a graph neural network (GNN) framework, CAMP enables physically motivated, systematically improvable potentials.
  The model demonstrates excellent performance across diverse systems, including periodic structures, small organic molecules, and two-dimensional materials, achieving accuracy, efficiency, and stability in molecular dynamics simulations that rival or surpass current leading models.
  CAMP provides a powerful tool for atomistic simulations to accelerate materials understanding and discovery.

 \end{abstract}

\maketitle

Machine learning interatomic potentials (MLIPs) have been widely employed in molecular simulations to investigate the properties of all kinds of materials, ranging from small organic molecules to inorganic crystals and biological systems.
MLIPs offer a balance between accuracy and efficiency~\cite{huang2021initio, unke2021machine, deringer2021gaussian, wen2021kliff}, making them a suitable choice between first-principles methods such as density functional theory (DFT) and classical interatomic potentials such as the embedded atom method (EAM)~\cite{daw1984embedded}.

Since the introduction of the pioneering Behler--Parrinello neural network (BPNN) potential~\cite{behler2007generalized} and the Gaussian approximation potential (GAP)~\cite{bartok2010gaussian}, many valuable MLIPs have been developed~\cite{chmiela2017machine,
zhang2018deep,
schutt2018schnet,
zubatyuk2019accurate,
gasteiger2020dimenet,
schutt2021equivariant,
ko2021fourth,
unke2021spookynet,
batatia2022mace,
batzner2022e,
takamoto2022teanet,
simeon2023tensornet}.
At the same time, techniques have emerged to assess the reliability of MLIP predictions by quantifying uncertainties and evaluating their influence on downstream properties~\cite{wen2020uncertainty, zhu2023fast, tan2023single, vita2024ltau, dai2024uncertainty}. Despite this progress, the pursuit of enhancing and developing new MLIPs remains ongoing.
The current leading MLIPs in terms of accuracy and stability in molecular dynamics (MD) simulations are those based on graph neural networks (GNNs).
These models represent an atomic structure as a graph and perform message passing between atoms to propagate information \cite{gilmer2017neural}.
The passed messages can be either scalars \cite{schutt2018schnet,gasteiger2020dimenet}, vectors \cite{schutt2021equivariant}, or, more generally, tensors \cite{batzner2022e, takamoto2022teanet, simeon2023tensornet, batatia2022mace}.
Using GNNs, a couple of universal MLIPs have recently been developed \cite{chen2022a, takamoto2022towards, deng2023chgnet, batatia2023a, xie2024gptff, zhang2023dpa, yang2024mattersim}, which can cover a wide range of materials with chemical species across the periodic table.

At the core of any MLIP lies a description of the atomic environment, which converts the information contained in an atomic neighborhood to a numerical representation that can then be updated and used in an ML regression algorithm.
The representation must satisfy certain symmetries (invariance to permutation of atoms, and invariance or equivariance to translation, rotation, and inversion) inherent to atomic systems.
Despite the diverse choice of ML regression algorithms, the atomic environment representation is typically constructed by expanding atomic positions on some basis functions.
Early works employ manually crafted basis functions for the bond distances and bond angles, such as the atom-centered symmetry functions used in BPNN \cite{behler2007generalized} and ANI-1 \cite{smith2017ani}.
More systematic representations have been developed and adopted in GAP \cite{bartok2010gaussian}, SNAP \cite{thompson2015spectral}, ACE \cite{drautz2019atomic}, NequIP \cite{batzner2022e}, and MACE \cite{batatia2022mace}, among others.
In these models, a representation is obtained by first expanding the atomic environment using a radial basis (e.g., Chebyshev polynomials) and the spherical harmonics angular basis to obtain tensorial equivariant features.
The equivariant features are then updated and finally contracted to yield the scalar potential energy.

Another approach seeks to develop atomic representations and MLIPs entirely in Cartesian space, bypassing the need for spherical harmonics.
This can potentially offer greater simplicity and reduced complexity compared to models using spherical representations.
The MTP model by Shapeev \cite{shapeev2016moment} adopted this approach, which constructs Cartesian moment tensors to capture the angular information of an atomic environment.
It has been shown that MTP and ACE are highly related given that a moment tensor can be expressed using spherical harmonics~\cite{dusson2022atomic}.
TeaNet~\cite{takamoto2022teanet} and TensorNet~\cite{simeon2023tensornet} have explored the use of Cartesian tensors limited to the second rank to create MLIPs.
The recent CACE model \cite{cheng2024cartesian} has extended the atomic cluster expansion to Cartesian space and subsequently built MLIPs on this foundation.
These works represent excellent efforts toward building MLIPs using a fully Cartesian representation of atomic environments. However, their performance has not yet matched that of models built using spherical tensors.

In this work, we propose the Cartesian Atomic Moment Potential (\model), a systematically improvable MLIP developed in Cartesian space under the GNN framework.
Inspired by MTP~\cite{shapeev2016moment}, \model employs physically motivated moment tensors to characterize atomic environments.
It then generates hyper moments through tensor products of these moment tensors, effectively capturing higher-order interactions.
We have devised specific construction rules for atomic and hyper moment tensors that direct message flow from higher-rank to lower-rank tensors, ultimately to scalars.
This approach aligns with our objective of modeling potential energy (a scalar quantity) and significantly reduces the number of moment tensors, enhancing computational efficiency.
These moment tensors are subsequently integrated into a message-passing GNN framework, undergoing iterative updates and refinement to form the final \model.
To evaluate the performance of the model, we have conducted tests across diverse material systems, including periodic LiPS inorganic crystals~\cite{batzner2022e} and bulk water~\cite{cheng2019ab}, non-periodic small organic molecules~\cite{chmiela2017machine}, and partially periodic two-dimensional (2D) graphene~\cite{wen2019hybrid}.
These comprehensive benchmark studies demonstrate that \model is an accurate, efficient, and stable MLIP for molecular simulations.

\section*{Results}
\label{sec:results}

\subsection*{Model Architecture}
\label{sec:model}

\model is a GNN model that processes atomic structures, utilizing atomic coordinates, atomic numbers, and cell vectors as input, and predicts the total potential energy, stresses on the structure, and the forces acting on individual atoms.

\begin{figure*}[bth!]
  \centering
  \includegraphics[width=0.8\linewidth]{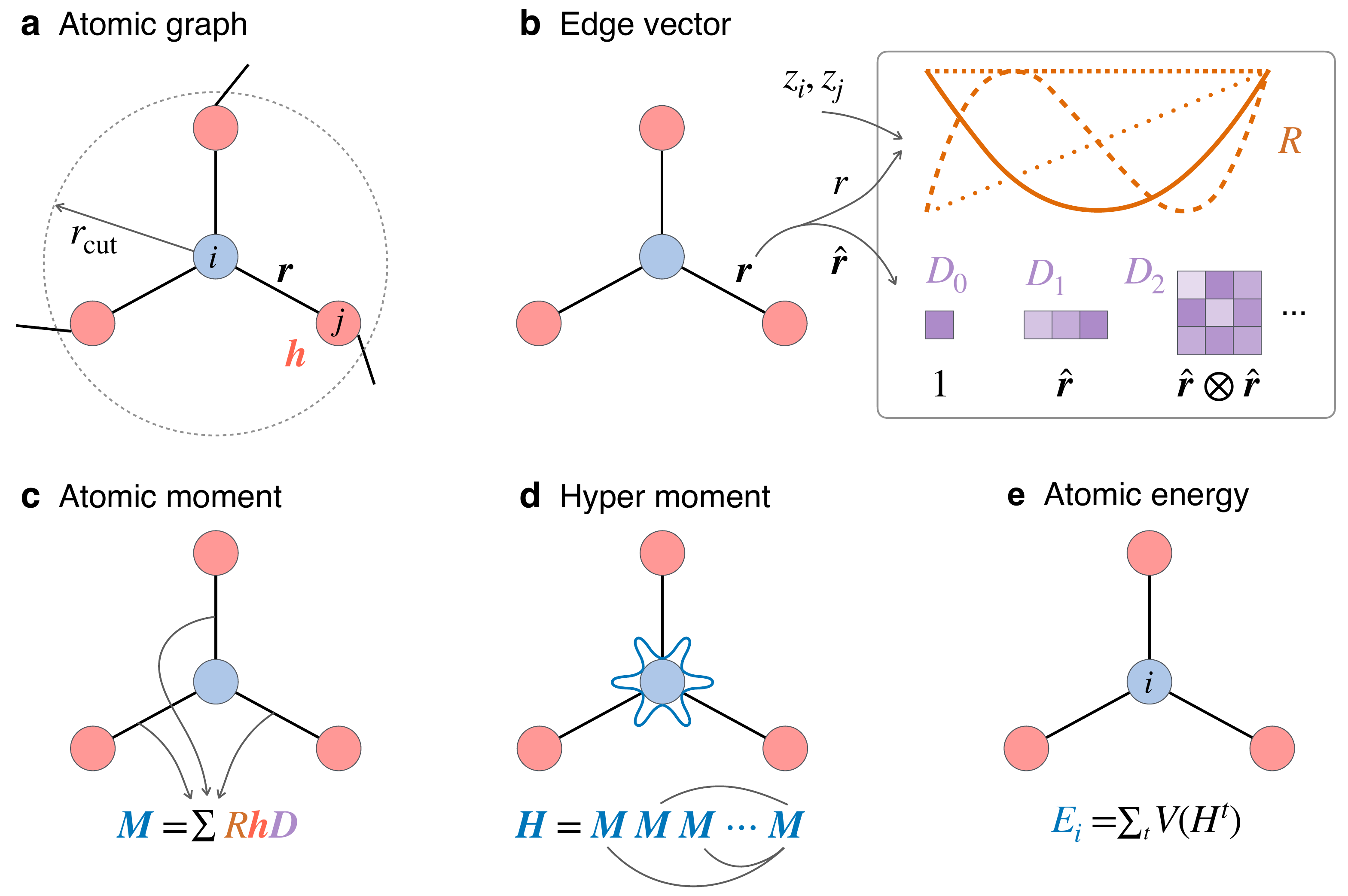}
  \caption{\textbf{Schematic overview of the \model model architecture.}
    \textbf{a} An atomic structure is represented as a graph.
    Each atom is associated with a feature $\bm h$, and all atoms within a distance of $r_\text{cut}$ to a center atom $i$ constitute its local atomic neighborhood.
    \textbf{b} Atomic numbers  $z_i$ and $z_j$ of a pair of atoms and their relative position vector $\bm r$ are expanded into the radial part $\bm R$ and angular part $\bm D$.
    \textbf{c} Atomic moment $\bm M$ of the central atom is obtained by aggregating information from neighboring atoms.
    \textbf{d} Hyper moment $\bm H$ of the central atom is computed by self-interactions between atomic moments.
    \textbf{e} Hyper moments $\bm H^t$ of different layers $t$ are used to construct the energy of the central atom via some mapping function $V$.
  }
  \label{fig:model:arch}
\end{figure*}

\fref{fig:model:arch} presents a schematic overview of the model architecture.
In the following, we discuss the key components and model design choices.

\textbf{Atomic graph.}
An atomic structure is represented as a graph $G(V, E)$ consisting of a set of nodes $V$ and a set of edges $E$.
Each node in $V$ represents an atom, and an edge in $E$ is created between two atoms if they are within a cutoff distance of $r_\mathrm{cut}$ (\fref{fig:model:arch}a).
An atom node $i$ is described by a 3-tuple $(\bm r^i, z^i, \bm h^i)$, where $\bm r^i$ is the position of the atom, $z^i$ is the atomic number, and $\bm h^i$ is the atom feature.
The vector $\bm r^{ij} = \bm r^j - \bm r^i$ associated with the edge from atom $i$ to atom $j$ gives their relative position.

The atom feature $\bm h^i$ is a set of tensors that carry two indices $u$ and $v$, becoming $\bm h^i_{uv}$ in its full form, where $u$ and $v$ denote the \emph{channel} and \emph{rank} of each feature tensor, respectively.
The use of channels allows for multiple copies of tensors of the same rank, helping to increase the expressiveness of the feature.

The atom features are mixed across different channels using a linear mapping:
\begin{equation}
  \bm h^i_{uv} = \sum_{u'} W_{uu'} \bm h^i_{u'v},
\end{equation}
where $W_{uu'}$ are trainable parameters.

\textbf{Radial basis.}
The edge length $r^{ij} = \|\bm r^{ij}\|$ is expanded using a set of radial basis functions, $B_u(r^{ij}, z^i, z^j)$, indexed by the channel $u$.
Following MTP \cite{shapeev2016moment, novikov2020mlip},
each basis function is a linear combination of the Chebyshev polynomials of the first kind $Q^\beta$:
\begin{equation} \label{eq:radial:basis}
  B_u(r^{ij}, z^i, z^j) = \sum_{\beta=0}^{N_\beta} W_{u z^i z^j}^\beta Q^\beta \left(\frac{r^{ij}}{r_\mathrm{cut}} \right),
\end{equation}
where $N_\beta$ denotes the maximum degree of the Chebyshev polynomials.
Separate trainable parameters $W_{u z^i z^j}^\beta$ are used for different atom pairs, enabling customized radial basis dependent on their atomic numbers $z^i$ and $z^j$.
Moreover, the radial basis $B_u$ allows the model to scale to any number of chemical species without increasing the model size.

\textbf{Angular part.}
The angular information of an atomic environment is contained in the normalized edge vectors, $\hat{\bm r} = \bm r / r$, where we have omitted the atom indices $i$ and $j$ in $\bm r^{ij}$ and $r^{ij}$ for simplicity.
Unlike many existing models that expand it on spherical harmonics, we directly adopt the Cartesian basis.
The rank-$v$ \emph{polyadic tensor} of $\hat{\bm r}$ is constructed as
\begin{equation} \label{eq:dyadic}
  \bm D_{v} = \hat{\bm r} \otimes \hat{\bm r} \otimes \dots \otimes \hat{\bm r
  }\quad (v\, \mathrm{of}\, \hat{\bm r}),
\end{equation}
where $\otimes$ denotes a tensor product.
For example, $\bm D_0 = 1$ is a scalar, $\bm D_1 = \hat{\bm r} = [\hat r_x, \hat r_y, \hat r_z]$ is a vector, and $\bm D_2 = \hat{\bm r} \otimes \hat{\bm r} $ is a rank-2 tensor, which can be written in a matrix form as
\begin{equation}
  \bm D_2 =
  \begin{bmatrix}
    \hat r_x^2       & \hat r_x \hat r_y & \hat r_x \hat r_z \\
    \hat r_y\hat r_x & \hat r_y^2        & \hat r_y \hat r_z \\
    \hat r_z\hat r_x & \hat r_z \hat r_y & \hat r_z^2
  \end{bmatrix},
\end{equation}
where $\hat r_x$, $\hat r_y$, and $\hat r_z$ are the Cartesian components of $\hat{\bm r}$ (\fref{fig:model:arch}b).
Any $\bm D_v$ is symmetric, as can be seen from the definition in \eref{eq:dyadic}.
In \model, the angular information of an atomic environment is captured by $\bm D_0$, $\bm D_1$, $\bm D_2$, and so on, which are straightforward to evaluate.
In full notation with atom indices, a polyadic tensor is written as $\bm D^{ij}_v$.

\textbf{Atomic moment.}
A representation of the local atomic environment of an atom is constructed from the radial basis, angular part, and atom feature, which we call the \emph{atomic moment}:
\begin{equation} \label{eq:atomic:moment}
  \bm M^i_{uv, p} = \sum_{j \in \mathcal{N}_i} R_{uvv_1v_2} \bm h^j_{uv_1} \contract \bm D^{ij}_{v_2},
\end{equation}
where $\mathcal{N}_i$ denotes the set of neighboring atoms within a distance of $r_\mathrm{cut}$ to atom $i$.
The radial part is obtained by passing the radial basis through a multilayer perceptron (MLP), $R_{uvv_1v_2} = \mathrm{MLP} (B_u)$.
For different combinations of $v$, $v_1$, and $v_2$, different MLPs are used.
The symbol $\contract$ denotes a degree-$c$ contraction between two tensors.
For example, for tensors $\bm A$ and $\bm B$ of rank-2 and rank-3, respectively, $\bm A \contractc{2} \bm B$ means $C_k = \sum_{ij} A_{ij}B_{ijk}$, resulting in a rank-1 tensor.
In other words, $c$ is the number of indices contracted away in each of the two tensors involved in the contraction; thus, the output tensor has a rank of $v = v_1 + v_2 - 2c$.
We denote the combination of $v_1$, $v_2$, and $c$ as a \emph{path}: $p = (v_1, v_2, c)$.
In \eref{eq:atomic:moment}, multiple paths can result in output tensors of the same rank $v$.
For example, $p = (2, 3, 2)$ and $p = (1, 2, 1)$ both lead to atomic moment rank $v=1$.
This is why the index $p$ is used in $\bm M_{uv,p}^i$ to denote the path from which the atomic moment is obtained.

The atomic moment carries significant physical implications.
For a physical quantity $Q$ at a distance $r$ from a reference point, the $n$th moment of $Q$ is defined as $r^n Q$.
For example, when $Q$ represents a force and $n=1$, $r Q$ corresponds to the torque.
\eref{eq:atomic:moment} generalizes this concept, with the feature $\bm h_{uv_1}^j$ of atom $j$ acting as $Q$, and $\bm D^{ij}_{v_2}$ serving as the ``distance'' $r$.
The radial part $R_{uvv_1v_2}$ functions as a weighting factor, modulating the relative contribution of different atoms $j$.
This is why we refer to the output in \eref{eq:atomic:moment} as the atomic moment.

By construction, the atomic moment is symmetric.
This is because, first, both $\bm h^j_{uv_1}$ (explained below around \eref{eq:feature:update}) and $\bm D^{ij}_{v_2}$ are symmetric;
second, an additional constraint $c=v_1<v_2$ is imposed.
The constraint means that the indices of $\bm h^j_{uv_1}$ and thus the output atomic moment has $v = v_2 - v_1$ indices from $\bm D^{ij}_{v_2}$.
A detailed example of evaluating \eref{eq:atomic:moment} is provided in the \sifull (\sishort).

Atomic moment tensors of the same rank $v$ from different paths are combined using a linear layer:
\begin{equation}
  \label{eq:path:mix}
  \bm M_{uv} = \sum_p W_{uv,p} \bm M_{uv,p},
\end{equation}
in which $W_{uv,p}$ are trainable parameters and the atom index $i$ is omitted for simplicity.

\textbf{Hyper moment.}
From atomic moments, we create the \emph{hyper moment} as
\begin{equation} \label{eq:hyper:moment}
  \bm H_{uv,p} = \bm M_{uv_1} \contractc{c_1} \bm M_{uv_2} \contractc{c_2} ... \contractc{c_{n-1}} \bm M_{uv_n}.
\end{equation}

Atomic moments capture two-body interactions between a center atom and its neighbors.
Three-body, four-body, and higher-body interactions are incorporated into the hyper moments via the self-interactions between atomic moments in \eref{eq:hyper:moment}.
The hyper moments can provide a complete description of the local atomic environment by increasing the order of interactions~\cite{shapeev2016moment,dusson2022atomic}, which is crucial for constructing systematically improvable MLIPs.
The hyper moments are analogous to the B-basis used in ACE~\cite{drautz2019atomic} and MACE~\cite{batatia2022mace}.

The primary target of an MLIP model is the potential energy, a scalar quantity.
Aligned with this, we construct hyper moments such that information flows from higher-rank atomic moment tensors to lower-rank ones, ultimately to scalars.
Specifically, we impose the following rule: let $\bm M_{uv_n}$ be the tensor of the highest rank among all atomic moment tensors in \eref{eq:hyper:moment}, then possible contractions are only between it and the other atomic moments, but not between $\bm M_{uv_1}$ and $\bm M_{uv_2}$, and so forth.
This design choice substantially reduces the number of hyper moments, which simplifies both the model architecture and the overall training process.
Moreover, similar to the atomic moment, the indices of lower-rank atomic tensors $\bm M_{uv_1}, \bm M_{uv_2}, ...$ are completely contracted away, and therefore we have $v = v_n - (v_1 + v_2 + ... + v_{n-1})$.
This guarantees that $\bm H_{uv,p}$ is constructed to be symmetric.
Multiple contractions can result in hyper moments of the same rank $v$; they are indexed by $p$.
A detailed example of evaluating \eref{eq:hyper:moment} to create hyper moments is provided in the \sishort.

\textbf{Message Passing.}
The messages to an atom $i$ from all neighboring atoms are chosen to be a linear expansion over the hyper moments:
\begin{equation}
  \bm m_{uv} = \sum_p W_{uv,p} \bm H_{uv,p}.
\end{equation}
Atom features are then updated using a residual connection \cite{he2015deep} by linearly combining the message and the atom feature of the previous layer:
\begin{equation} \label{eq:feature:update}
  \bm h^t_{uv} = \sum_{u'} W_{1,uu'} \bm m_{u'v}
  + \sum_{u'} W_{2,uu'} \bm h^{t-1}_{u'v},
\end{equation}
where $t$ is the layer index.
Atom features in the input layer $\bm h_{uv'}^0$ consist only of scalars $h_{u0}^0$, obtained as a $u$-dimensional embedding of the atomic number $z$.
By construction, $\bm h^t_{uv}$ is symmetric.

\textbf{Output.}
The message passing is performed multiple times $T$.
We find that typical values of two or three are good enough based on benchmark studies to be discussed in the following sections.
The atomic energy of atom $i$ is then obtained from the scalar atom features $h^{i,t}_{u0}$ from all layers:
\begin{equation}
  E^i = \sum_{t=1}^T V(\bm h^{i,t}_{u0}).
\end{equation}
$V$ is set to an MLP for the last layer $T$, and a linear function for others layers, that is,
$V(\bm h^{i,t}_{u0}) = \sum_u W^t_u \bm h^{i,t}_{u0}$ for $t<T$.
Atomic energies of all atoms are summed to get the total potential energy
\begin{equation}
  E = \sum_i E^i.
\end{equation}
Forces on atom $i$ can then be computed as
\begin{equation}
  \bm{F}_i = - \frac{\partial E}{\partial \bm r_i}.
\end{equation}

\textbf{Computational complexity.}
The most computationally demanding part of \model is creating the atomic moments and hyper moments.
Let $v_\text{max}$ represent the maximum rank of tensors used in the model.
This implies that all features $\bm h$, atomic moments $\bm M$, and hyper moments $\bm H$ have ranks up to $v_\text{max}$.
The time complexity for constructing both the atomic moment in \eref{eq:atomic:moment} and the hyper moment in \eref{eq:hyper:moment} is $\mathcal{O}(3^{v_\text{max}})$ (detailed analysis in the \sishort).
In contrast, for models based on spherical tensors (such as NequIP \cite{batzner2022e} and MACE \cite{batatia2022mace}), the time complexity of the Clebsch--Gordan tensor product is $\mathcal{O}(L^6)$, where $L$ is the maximum degree of the irreducible spherical tensors.
A Cartesian tensor (e.g., $\bm M$ and  $\bm H$) of rank $v$ can be decomposed as the sum of spherical tensors of degrees up to $v$ \cite{zee2016group}; therefore, as far as the computational cost is concerned, $v_\text{max}$ and $L$ can be thought to be equivalent to each other.
Asymptotically, as $L$ (or $v_\text{max}$) increases, the exponential-scaling Cartesian models are computationally less efficient than the polynomial-scaling spherical models.
Nevertheless, empirical evidence suggests that, for many material systems, the adopted values of $L$ or ($v_\text{max}$) are below five \cite{batzner2022e, batatia2022mace, wen2024an}.
In such cases, Cartesian models should be more computationally efficient.
We note that there exist techniques to reduce the computational cost of the Clebsch--Gordan tensor product in spherical models from $\mathcal{O}(L^6)$ to $\mathcal{O}(L^3)$, such as reducing SO(3) convolutions to SO(2) \cite{passaro2023reducing} and replacing SO(3) convolutions by matrix multiplications \cite{Maennel2024Dec}.
When such techniques are adopted, spherical models can be more efficient.

\subsection*{Inorganic crystals}

\begin{table}[tb]
  \caption{\textbf{Performance of \model on the LiPS dataset}.
    MAEs of energy and forces are in the units of meV/atom and meV/\AA, respectively.
    Errors shown in parentheses are obtained as the standard deviations from five independently trained \model models.
    }
  \label{tab:lips}
  \begin{tabular}{ccccccc}
    \hline
    Training size &        & NequIP        & \model        \\
    \hline
    10            & Energy & 2.03          & \textbf{1.63} (0.16) \\
                  & Forces & 97.8          & \textbf{76.4} (3.3) \\
    100           & Energy & \textbf{0.44}          & \textbf{0.44} (0.02) \\
                  & Forces & 25.8          & \textbf{21.6} (0.6) \\
    1000          & Energy & \textbf{0.12} & \textbf{0.12} (0.00) \\
                  & Forces & 7.7           & \textbf{7.4} (0.4)  \\
    2500          & Energy & \textbf{0.08} & \textbf{0.08} (0.00)  \\
                  & Forces & \textbf{4.7}  & 4.9 (0.1)           \\
    \hline
  \end{tabular}
\end{table}

\begin{figure}[tb]
  \centering
  \includegraphics[width=0.9\linewidth]{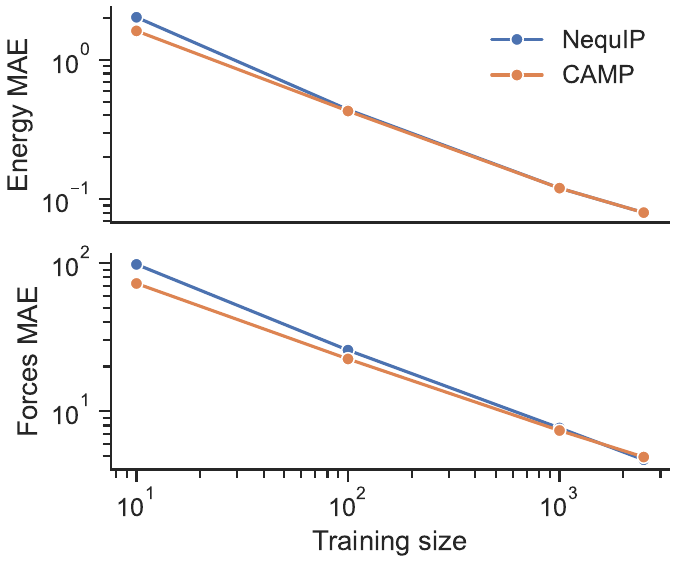}
  \caption{\textbf{Learning curve of \model.}
    On a log-log scale, the MAEs of both energy (meV/atom) and forces (meV/\AA) decrease linearly with the training set size.
  }
  \label{fig:learning:curve}
\end{figure}

\begin{table*}[tbh]
   \footnotesize
  \caption{\textbf{Results on LiPS for models developed with 19000 training samples}.
    MAE of forces and MAE of \ce{Li+} diffusivity are reported in meV/\AA\ and \SI{e-5}{\cm^2\per\s}, respectively.
    The reference diffusivity is \SI{1.35e-5}{\cm^2\per\s}.
    Stability is measured by MD simulation time in picoseconds, with standard deviations obtained from five independent runs shown in parentheses.
    All results are obtained from \olcite{fu2023forces}, except those for \model.
  }
  \label{tab:lips2}
  \begin{tabular}{ccccccccccc}
    \hline
                & DeepPot-SE \cite{deeppotse}
                & SchNet \cite{schutt2018schnet}
                & DimeNet \cite{gasteiger2020dimenet}
                & PaiNN \cite{schutt2021equivariant}
                & SphereNet \cite{liu2022spherical}
                & ForceNet \cite{hu2021forcenet}
                & GemNet-T \cite{gemnet}
                & NequIP    \cite{batzner2022e}
                & \model                                                                                                          \\
    \hline
    Forces      & 40.5                              & 28.8       & 3.2        & 11.7       & 8.3        & 12.8       & 1.3        
                & 3.7
                & 3.6                                                                                                             \\
    Diffusivity &                                   & 0.38       & 0.30       & 0.40       & 0.40       &            & 0.24       
                & 0.34                              & 0.18                                                                        \\
    Stability   & 4 (3)                         & 50 (0) & 48 (4) & 50 (0) & 50 (0) & 26 (8) & 50 (0) 
                & 50 (0)                        & 50 (0)                                                                  \\
    \hline
  \end{tabular}
\end{table*}

\begin{figure*}[htb]
  \centering
  \includegraphics[width=1.\linewidth]{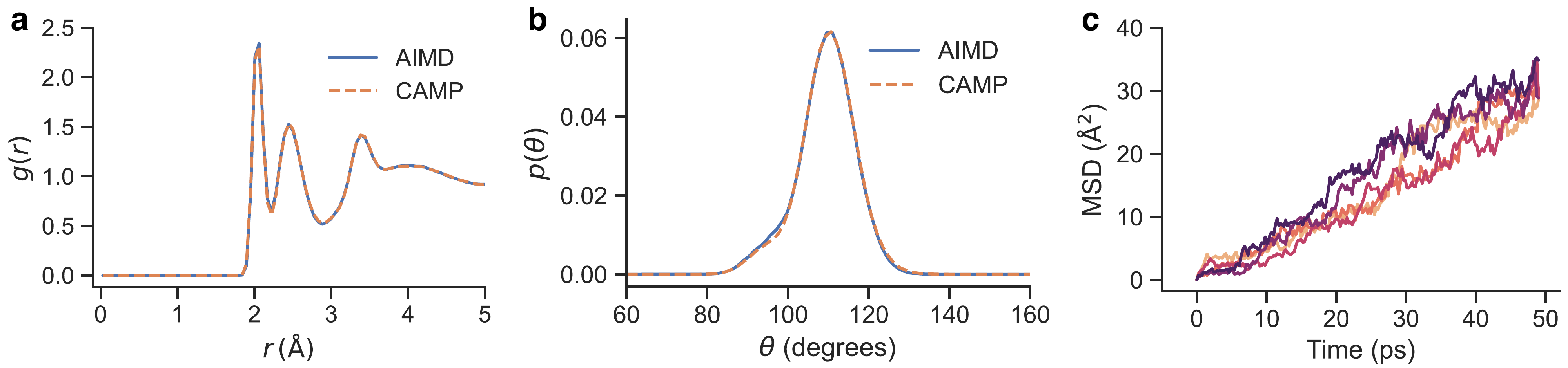}
  \caption{\textbf{MD simulation results on LiPS}.
    \textbf{a} Radial distribution function.
    \textbf{b} Angular distribution function for the S--P--S tetrahedral angle.
    \textbf{c} Mean squared displacement of \ce{Li+} in LiPS.
    Five MD simulations are performed, and
    the calculated \ce{Li+} diffusion coefficient is  $D = (1.08\pm0.08) \times10^{-5}$~\si{\cm^2\per\s}.
  }
  \label{fig:lips:results}
\end{figure*}

We first test \model on a dataset of inorganic lithium phosphorus sulfide (LiPS) solid-state electrolytes~\cite{batzner2022e}.
Multiple models were trained using varying numbers of training samples (10, 100, 1000, and 2500), validated on 1000 samples, and tested on 5000 samples (training details in \method{sec:methods}).
Mean absolute errors (MAEs) of energy and forces on the test set are listed in \tref{tab:lips}.
Compared with NequIP \cite{batzner2022e}, \model yields smaller or equal errors on seven of the eight tasks.
The learning curve in \fref{fig:learning:curve} exhibits a linear decrease in both energy and forces MAEs on a log-log scale as the number of training samples increases.
These results demonstrate \model's accuracy and its capacity for further improvement with increased data.

We further examine the performance in MD simulations, focusing on computing the diffusivity of lithium ions (\ce{Li+}) in LiPS.
Diffusivity is a crucial characteristic for assessing the potential of solid materials as electrolytes in next-generation solid-state batteries.
Following existing benchmark studies~\cite{fu2023forces}, we trained another model with 19000 structures with the same validation and test sets discussed above.
Five structures were randomly selected from the test set and an independent MD simulation was performed for each structure.
As seen in \fref{fig:lips:results}, \model accurately captures the structural information of LiPS, reproducing the radial distribution function (RDF) and the angular distribution function (ADF) of the S--P--S tetrahedral angle (Fig.~S1 in the \sishort) from ab initio molecular dynamics (AIMD) simulations.

\model can yield stable MD simulations.
It is known that MLIPs that perform well on energy and forces do not necessarily guarantee high-quality MD simulations.
In particular, MD simulations can collapse due to model instability.
Therefore, before calculating the diffusivity, we checked the simulation stability.
Stability is measured as the difference between the RDF averaged over a time window and the RDF of the entire simulation (see \method{sec:methods}).
\model shows excellent stability, reaching the entire simulation time of \SI{50}{\ps} in all five simulations (\tref{tab:lips2}).
In \olcite{fu2023forces}, a small timestep of \SI{0.25}{\fs} was adopted in the MD simulations using the models in \tref{tab:lips2}.
We also tested \model with a large timestep of \SI{1}{\fs} and found that all five simulations are still stable up to \SI{50}{\ps} (see Fig.~S2 in the \sishort).

The diffusion coefficient $D$ of \ce{Li+} in LiPS calculated from \model agrees well with the AIMD result.
After confirming the stability, $D$ was calculated using the Einstein equation by fitting a linear line between the mean squared displacement (MSD) of \ce{Li+} and the correlation time (see \method{sec:methods}).
The MSDs of the five MD runs are shown in \fref{fig:lips:results}c, from which the diffusion coefficient is calculated as $D = (1.08\pm0.08) \times10^{-5}$~\si{\cm^2\per\s}.
This agrees reasonably well with AIMD result of \SI{1.37e-5}{\cm^2\per\s}~\cite{batzner2022e}.

\subsection*{Bulk water}

\begin{figure}[hbt]
  \includegraphics[width=0.9\linewidth]{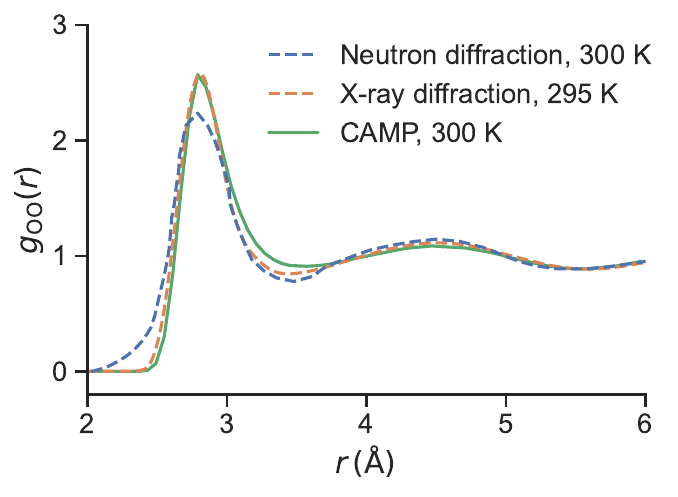}
  \caption{\textbf{Oxygen--Oxygen RDF of water}.
    Experimental neutron diffraction \cite{soper1997site} and x-ray diffraction \cite{skinner2014the} results are shown for comparison.
  }
  \label{fig:water:rdf}
\end{figure}

\begin{table*}[htb!]
  \caption{\textbf{Model performance on the water dataset}.
    RMSEs of energy and forces are in the units of meV/atom and meV/\AA, respectively.
    Speed is measured as completed steps per second in MD simulations of the water system of 192 atoms, using a single NVIDIA V100 GPU.
    All energy and forces RMSE results, except \model, are from \olcite{cheng2024cartesian}.
    DeePMD and NequIP running speed results are from \olcite{fu2023forces}.
    Errors shown in parentheses are obtained as the standard deviations from five independently trained \model models.
  }
  \label{tab:water}
  \begin{tabular}{cccccccccc}
    \hline
           & BPNN \cite{cheng2019ab}
           & ACE \cite{drautz2019atomic}
           & REANN \cite{zhang2021physically}
           & DeePMD \cite{zhang2018deep}
           & NequIP \cite{batzner2022e}
           & MACE \cite{batatia2022mace}
           & CACE \cite{cheng2024cartesian}
           & \model                                                                                            \\
    \hline
    Energy & 2.3                              & 1.7 & 0.8 & 2.1  & 0.94 & 0.63 & \textbf{0.59} & \textbf{0.59} (0.002) \\
    Forces & 120                              & 99  & 53  & 92   & 45   & 36   & 47            & \textbf{34} (0.1)   \\
    Speed  &                                  &     &     & 61.8 & 3.9  &      & 9.8           & 14.2          \\
    \hline
  \end{tabular}
\end{table*}

To evaluate \model's ability to model complex liquid systems, we test it on a dataset of bulk water~\cite{cheng2019ab}.
See \method{sec:methods} Dataset and training details are in
Root-mean-square errors (RMSEs) of energy and forces on the test set are listed in \tref{tab:water}.
In general, GNN models (REANN \cite{zhang2021physically}, NequIP \cite{batzner2022e}, MACE \cite{batatia2022mace}, and CACE \cite{cheng2024cartesian}) have smaller errors than single-layer ACE \cite{drautz2019atomic} and descriptors-based BPNN \cite{cheng2019ab}.
Our \model achieves the best performance on both energy and forces, with RMSEs of 0.59~meV/atom and 34~meV/\AA, respectively.
We observe that CACE \cite{cheng2024cartesian} demonstrates comparable RMSEs in energy when compared to \model.
CACE \cite{cheng2024cartesian} is a recent model also developed entirely in Cartesian space.
Similar to \model, it also first constructs the atomic basis (\eref{eq:atomic:moment}), then builds the product basis on top of the atomic basis (\eref{eq:hyper:moment}), and iteratively updates the features under a GNN framework to refine the representations.
A major difference is that CACE is formulated based on the atomic cluster expansion, while \model is designed using the atomic moment.
The connections between \model and CACE, as well as other models, are further elaborated in \discussion{sec:discussion}.

\model accurately reproduces experimental structural and dynamical properties of water.
\fref{fig:water:rdf} shows the oxygen--oxygen RDF of water from MD simulations using \model (computation details in \method{sec:methods}), together with experimental observations obtained from neutron diffraction~\cite{soper1997site} and x-ray diffraction~\cite{skinner2014the}.
The RDF by \model almost overlaps with the x-ray diffraction result, demonstrating its ability to capture the delicate structural features of water.
We also calculated the diffusion coefficient of water (\SI{1}{\gram\per\cm^3}) at 300~K to be $D=(2.79\pm0.18) \times 10^{-5}$~\si{\cm^2\per\s}, which is in excellent agreement with AIMD results of \SI{2.67e-5}{\cm^2\per\s}~\cite{marsalek2017quantum}.
The model shows superior stability for water, producing stable MD simulations at high temperatures up to 1500~K.
The RDF and the diffusion coefficient at various high temperatures are presented in Fig.~S3 in the \sishort.

We examine the efficiency of \model by measuring the number of completed steps per second in MD simulations.
It is not a surprise that DeePMD \cite{zhang2018deep} runs much faster due to its simplicity in model architecture, although its accuracy falls short when compared with more recent models (\tref{tab:water}).
For those more accurate models, \model is about 1.4 and 3.6 times faster than CACE \cite{cheng2024cartesian} and NequIP \cite{batzner2022e}, respectively, on the water system of 192 atoms.
More running speed data of other models are provided in Table~S1 in the \sishort.

\subsection*{Small organic molecules}

\begin{table*}[tbh]
  \caption{\textbf{Model performance on the MD17 dataset}.
    MAEs of energy and forces are in the units of meV and meV/\AA, respectively.
    \textbf{Bold}: smallest; \underline{underline}: second smallest.
    Errors shown in parentheses are obtained as the standard deviations from five independently trained \model models.
  }
  \label{tab:md17}
  \begin{tabular}{lcccccccccc}
    \hline
    Molecule       &
                   & SchNet \cite{schutt2018schnet}
                   & DimeNet \cite{dimnet}
                   & sGDML \cite{chmiela2017machine}
                   & PaiNN \cite{schutt2021equivariant}
                   & NewtonNet \cite{haghighatlari2022newtonnet}
                   & NequIP \cite{batzner2022e}
                   & \model                                                                                                                                     \\
    \hline
    Aspirin        & Energy                                      & 16.0 & 8.8  & 8.2  & \underline{6.9}  & 7.3             & \textbf{5.7}    & 8.7 (0.7)              \\
                   & Forces                                      & 58.5 & 21.6 & 29.5 & \underline{14.7} & 15.1            & \textbf{8.0}    & 20.5 (1.5)             \\
    Ethanol        & Energy                                      & 3.5  & 2.8  & 3.0  & 2.7              & 2.6             & \textbf{2.2}    & \underline{2.5} (0.1)  \\
                   & Forces                                      & 16.9 & 10.0 & 14.3 & 9.7              & 9.1             & \textbf{3.1}    & \underline{7.6} (0.4)  \\
    Malonaldehyde  & Energy                                      & 5.6  & 4.5  & 4.3  & \underline{3.9}  & 4.2             & \textbf{3.3}    & \underline{3.9} (0.0) \\
                   & Forces                                      & 28.6 & 16.6 & 17.8 & 13.8             & 14.0            & \textbf{5.6}    & \underline{12.7} (0.3) \\
    Naphthalene    & Energy                                      & 6.9  & 5.3  & 5.2  & 5.0              & 5.1             & \underline{4.9} & \textbf{4.2} (0.0)     \\
                   & Forces                                      & 25.2 & 9.3  & 4.8  & \underline{3.3}  & 3.6             & \textbf{1.7}    & 4.0 (0.3)              \\
    Salicylic acid & Energy                                      & 8.7  & 5.8  & 5.2  & \underline{4.9}              & 5.0             & \textbf{4.6}     & \textbf{4.6} (0.3)    \\
                   & Forces                                      & 36.9 & 16.2 & 12.1 & \underline{8.5}  & \underline{8.5} & \textbf{3.9}    & 11.0 (2.1)           \\
    Toluene        & Energy                                      & 5.2  & 4.4  & 4.3  & \underline{4.1}  & \underline{4.1} & \textbf{4.0}    & 4.2 (0.2)              \\
                   & Forces                                      & 24.7 & 9.4  & 6.1  & 4.1              & \underline{3.8} & \textbf{2.0}    & 4.6 (0.6)  \\
    Uracil         & Energy                                      & 6.1  & 5.0  & 4.8  & 4.5              & 4.6             & \underline{4.5} & \textbf{4.4} (0.0)     \\
                   & Forces                                      & 24.3 & 13.1 & 10.4 & \underline{6.0}              & 6.5 & \textbf{3.3}    & 7.2 (0.3)              \\
    \hline
  \end{tabular}
\end{table*}

The MD17 dataset consists of AIMD trajectories of several small organic molecules~\cite{chmiela2017machine}.
For each molecule, we trained \model on 950 random samples, validated on 50 samples, and the rest were used for testing (dataset and training details in \method{sec:methods}).
MAEs of energy and forces by \model and other MLIPs are listed in \tref{tab:md17}.
While NequIP \cite{batzner2022e} generally maintains high performance, \model demonstrates high competitiveness.
It surpasses NequIP in energy predictions for three out of seven molecules and ranks second in both energy and force predictions across various cases.

\begin{figure*}[bth]
  \centering
  \includegraphics[width=0.8\linewidth]{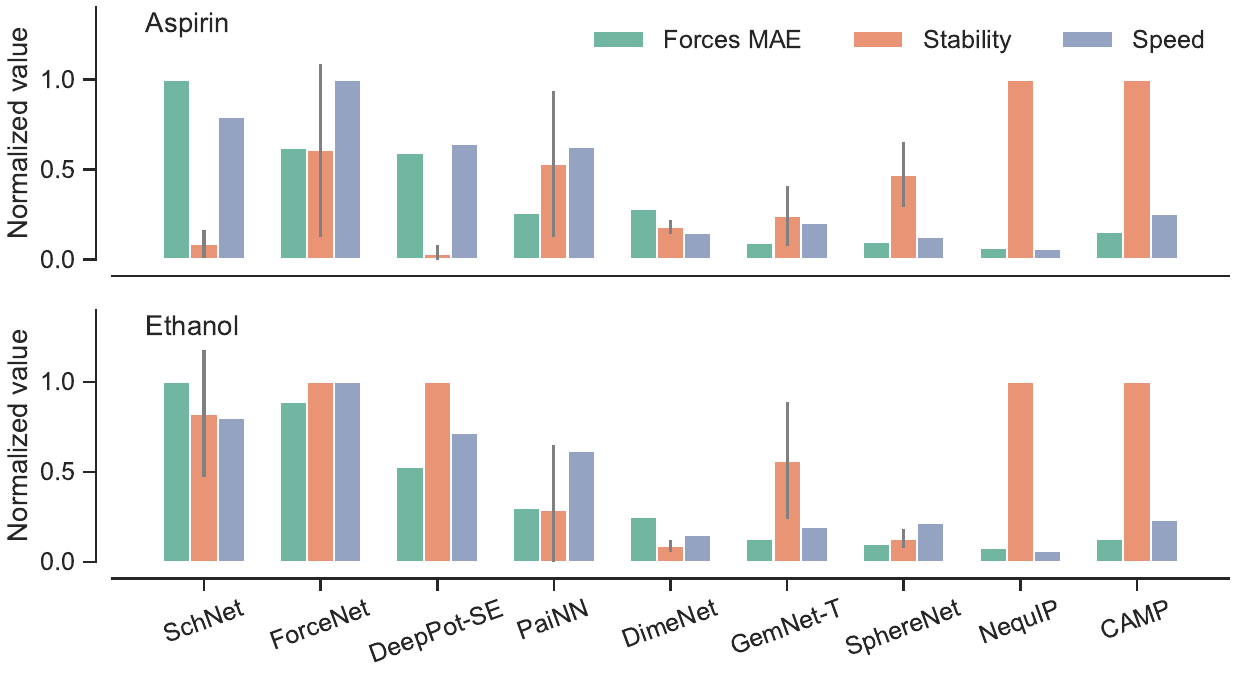}
  \caption{\textbf{Comparison of MAE of forces, MD stability, and running speed on small molecules.}
    The forces MAE, stability, and running speed are normalized by their maximum values for easier visualization.
    All numerical values are provided in Table~S2 in the \sishort.
    Error bars indicate the standard deviation from five MD runs using different starting atomic positions.
    Computational speed is measured on a single NVIDIA V100 GPU.
  }
  \label{fig:md17:hist}
\end{figure*}

We further tested the ability of \model to maintain long-time stable MD simulations.
Following the benchmark study in \olcite{fu2023forces}, 9500, 500, and 10000 molecules were randomly sampled for training, validation, and testing, respectively.
With the trained models, we performed MD simulations at \SI{500}{\kelvin} for \SI{300}{\ps} and examined the stability by measuring the change in bond length during the simulations (see \method{sec:methods}).
\fref{fig:md17:hist} presents the results for Aspirin and Ethanol (numerical values in Table~S2 in the \sishort).
In general, a small forces MAE does not guarantee stable MD simulations \cite{fu2023forces}, as seen from a comparison between ForceNet and GemNet-T.
GemNet-T has a much smaller MAE of forces, but its stability is not as good as ForceNet.
For \model, its MAE of forces is slightly larger than that of, e.g.\ GemNet-T and SphereNet, but it can maintain stable MD simulations for the entire simulation time of \SI{300}{\ps}, a significant improvement over most of the existing models.

In terms of running speed, similar patterns are observed here as those reported in the water system.
Models based on simple scalar features such as SchNet~\cite{schutt2018schnet} and DeepPot-SE~\cite{deeppotse} are faster than those based on tensorial features, although they are far less accurate (\fref{fig:md17:hist}).
For both Aspirin and Ethanol, \model can complete about 33 steps per second in MD simulations on a single NVIDIA V100 GPU, which is approximately 1.2, 1.4, and 3.9 times faster than GemNet-T~\cite{gemnet}, SphereNet~\cite{liu2022spherical}, and NequIP~\cite{batzner2022e}, respectively.

The MD17 dataset was adopted due to the excellent benchmark study by Fu et al.~\cite{fu2023forces}, which provides a consistent basis for comparing different models.
However, it has been observed that the MD17 dataset contains significant numerical errors, prompting the introduction of the rMD17 dataset to address this issue~\cite{christensen2020Oct}.
We evaluated \model on the aspirin and ethanol molecules from the rMD17 dataset, and it demonstrates competitive performance.
While \model has errors higher than those of leading spherical models like NequIP~\cite{batzner2022e}, Allegro~\cite{musaelian2023learning}, and MACE~\cite{batatia2022mace}, it outperforms models such as GAP~\cite{bartok2010gaussian}, FCHL~\cite{faber2018alchemical}, ACE~\cite{drautz2019atomic}, and PaiNN~\cite{schutt2021equivariant}.
Detailed results are presented in Table~S3 of the \sishort.

\subsection*{Two-dimensional materials}

\begin{table*}[hbt!]
  \caption{\textbf{Results on the bilayer graphene dataset}.
    MAEs of the energy and forces are in the units of meV/atom and meV/\AA, respectively.
    Errors shown in parentheses are obtained as the standard deviations from five independently trained \model models.
  }
  \label{tab:bilayer:graphene}
  \begin{tabular}{cccccc}
    \hline
           & AIREBO \cite{stuart2000airebo} & AIREBO-M \cite{o2015airebo} & LCBOP \cite{los2003intrinsic} & hNN \cite{wen2019hybrid} & \model \\
    \hline
    Energy & 659.9                          & 671.2                       & 718.3                         & 1.4                      & 0.35 (0.05)    \\
    Forces & 194.6                          & 186.9                       & 166.1                         & 46.0                     & 6.26 (0.04)    \\
    \hline
  \end{tabular}
\end{table*}

\begin{figure*}[tbh]
  \centering
  \includegraphics[width=0.9\textwidth]{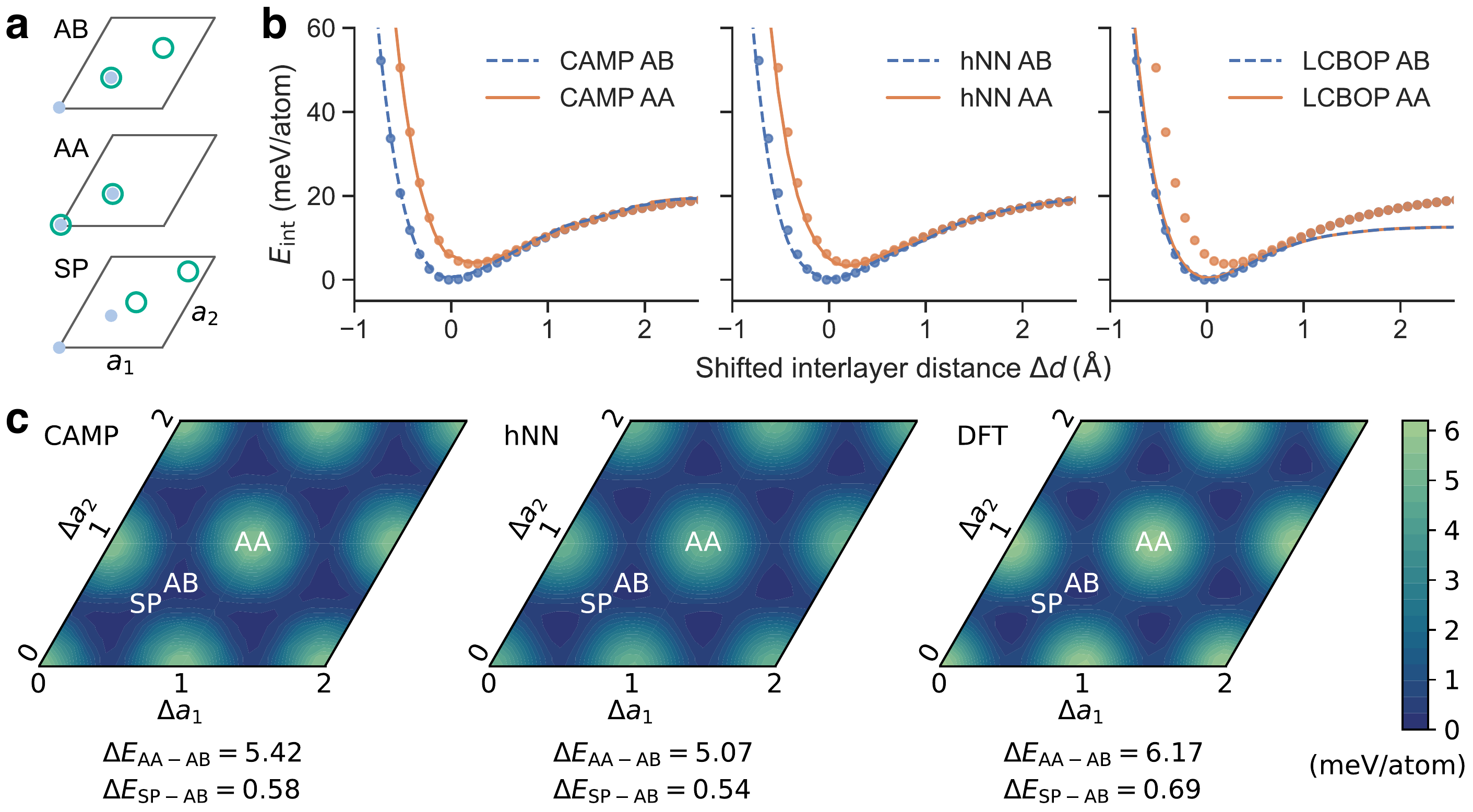}
  \caption{\textbf{Interlayer energetics of bilayer graphene}.
    \textbf{a} Bilayer graphene in AB, AA, and saddle point (SP) stacking, where the blue solid dots represent atoms in the bottom layer, and the green hollow dots represent atoms in the top layer.
    \textbf{b} Interlayer energy $E_\text{int}$ versus layer distance $d$ of bilayer graphene in AB and AA stacking. The dots represent DFT results.
    The layer distance is shifted such that $\Delta d = d - d_0$, where $d_0=\SI{3.4}{\angstrom}$ is the equilibrium layer distance.
    \textbf{c} Generalized stacking fault energy of bilayer graphene.
    This is obtained by sliding the top layer against the bottom layer at a fixed layer distance of $d_0$, where $\Delta a_1$ and $\Delta a_2$ indicate the shifts of the top layer against the bottom layer, along the lattice vectors $a_1$ and $a_2$, respectively.
  }
  \label{fig:bilayer:graphene}
\end{figure*}

In addition to periodic systems and small organic molecules, we evaluate the performance of \model on partially periodic 2D materials.
These materials exhibit distinct physical and chemical properties.
Despite their significance, to the best of our knowledge, no standardized benchmark dataset exists for MLIPs for 2D materials.
We have thus constructed a new DFT dataset of bilayer graphene, building upon our previous investigations of carbon systems \cite{wen2019hybrid}.
See \method{sec:methods} for detailed information on this new dataset.

Widely used empirical potentials for carbon systems, such as AIREBO \cite{stuart2000airebo}, AIREBO-M \cite{o2015airebo}, and LCBOP \cite{los2003intrinsic} have large errors in predicting both the energy and forces of bilayer graphene.
The MAEs on the test set against DFT results are on the order of hundreds of meV/atom for energy and meV/\AA\ for forces (\tref{tab:bilayer:graphene}).
This is not too surprising given that: first, these potentials are general-purpose models for carbon systems, which are not specifically designed for multilayer graphene systems; second, their training data were experimental properties and/or DFT calculations using different density functionals from the test set.
MLIPs such as hNN~\cite{wen2019hybrid} (a hybrid model that combines BPNN \cite{behler2007generalized} and Lennard--Jones \cite{lennard1931cohesion}), which was trained on the same data as used here,  can significantly reduce the errors to 1.4~meV/atom for energy and 46.0~meV/\AA\ for forces.
\model can further drive the MAEs down to 0.3~meV/atom for energy and 6.3~meV/\AA\ for forces.

It is interesting to investigate the interlayer interaction between the graphene layers, which controls many structural, mechanical, and electronic properties of 2D materials \cite{geim2013van, wen2018dihedral}.
Here, we focus on the energetics in different stacking configurations: AB, AA, and saddle point (SP) stackings (\fref{fig:bilayer:graphene}a).
Empirical models such as LCBOP cannot distinguish between the different stacking states at all.
The interlayer energy versus layer distance curves are almost identical between AB and AA (\fref{fig:bilayer:graphene}b), and the generalized stacking fault energy surface is nearly flat (not shown), with a maximum value on the order of 0.01~meV/atom.
This is also the case for AIREBO and AIREBO-M (we refer to \olcite{wen2019hybrid} for plots).
On the contrary, \model and hNN can clearly distinguish between the stackings.
Both the interlayer energy versus layer distance curves and the generalized stacking fault energy surface agree well with DFT references, and \model has a slightly better prediction of the energy barrier ($\Delta E_\mathrm{SP-AB}$) and the overall energy corrugation ($\Delta E_\mathrm{AA-AB}$).
The hNN model is specifically designed as an MLIP for 2D materials, and its training process is complex, requiring separate training of the Lennard--Jones and BPNN components.
In contrast, \model does not require such special treatment and is straightforward to train.

\section*{Discussion}
\label{sec:discussion}

In this work, we develop \model, a new class of MLIP that is based on atomic moment tensors and operates entirely in Cartesian space.
\model is designed to be physically inspired, flexible, and expressive, and it can be applied to a wide range of systems, including periodic structures, small organic molecules, and 2D materials.
Benchmark tests on these systems demonstrate that \model achieves performance surpassing or comparable to current leading models based on spherical tensors in terms of accuracy, efficiency, and stability.
It is robust and straightforward to train, without the need to tune a large number of hyperparameters.
In all the tested systems, we only need to set four hyperparameters: the number of channels $u$, maximum tensor rank $v_\text{max}$, number of layers $T$, and cutoff radius $r_c$ to achieve good performance.

\model is related to existing models in several ways.
As shown in \olcite{dusson2022atomic}, ACE \cite{drautz2019atomic} and MTP \cite{shapeev2016moment} can be viewed as the same model, with the former constructed in spherical space while the latter in Cartesian space.
Both are single-layer models without iterative feature updates.
Loosely speaking, MACE \cite{batatia2022mace} and \model can be regarded as a generalization of the spherical ACE and the Cartesian MTP, respectively, to multilayer GNNs with iterative feature updates and refinement.
The atomic moment in \eref{eq:atomic:moment} and hyper moment in \eref{eq:hyper:moment} are related to the A-basis and B-basis in ACE and MACE.
The recent CACE model is related to MACE and \model.
Both MACE and CACE are based on atomic cluster expansion, but MACE implements this expansion in spherical space while CACE builds it in Cartesian space.
Both CAMP and CACE develop features in Cartesian space, but CAMP uses atomic moments while CACE employs atomic cluster expansion.
Moreover, \model generalizes existing Cartesian tensor models such as TeaNet \cite{takamoto2022teanet} and TensorNet \cite{simeon2023tensornet} that use at most second-rank tensors to tensors of arbitrary rank.
Despite these connections, \model is unique in its design of the atomic moment and hyper moment, and the selection rules that govern the tensor contractions.
These characteristics make \model a physically inspired, flexible, and expressive model.

Beyond energy and forces, \model can be extended to model tensorial properties.
The atom features used in \model are symmetric Cartesian tensors, which can be used to output tensorial properties with slight modifications to the output block.
For example, NMR tensors can be modeled by selecting and summing the scalar, vector, and symmetric rank-2 tensor components of the hyper moments, rather than using only the scalar component for potential energies.
However, the current implementation of \model in PyTorch has certain limitations.
All feature tensors are stored in full form, without exploiting the fact that they are symmetric.
In addition, it is possible to extend \model to use irreducible representations of Cartesian tensors (basically symmetric traceless tensors), analogous to spherical irreducible representations used in spherical tensor models.
Leveraging these symmetries and irreducible representations could further enhance model accuracy and improve time and memory efficiency.
These are directions for future work.

\section*{Methods}
\label{sec:methods}

\textbf{Dataset}.
The LiPS dataset~\cite{batzner2022e} consists of 250001 structures of lithium phosphorus sulfide (\ce{Li_{6.75}P3S11}) obtained from an AIMD trajectory.
Each structure has 83 atoms, with 27 Li, 12 P, and 44 S.
The data are randomized before splitting into training, validation, and test sets.
The water dataset~\cite{cheng2019ab} consists of 1593 water structures, each with 192 atoms, generated with AIMD simulations at 300~K. 
The data are randomly split into training, validation, and test sets with a ratio of 90:5:5.
The MD17 dataset~\cite{chmiela2017machine} consists of AIMD trajectories of several small organic molecules.
The number of data points for each molecule ranges from 133000 to 993000.
Our new bilayer graphene dataset is derived from~\olcite{wen2019hybrid}, with 6178 bilayer graphene configurations from stressed structures, atomic perturbations, and AIMD trajectories.
The data were generated using DFT calculations with the PBE functional~\cite{perdew1996generalized}, and the
many-body dispersion correction method~\cite{tkatchenko2012accurate} was used to account for van der Waals interactions.
The data are split into training, validation, and test sets with a ratio of 8:1:1.

\textbf{Model training}.
The models are trained by minimizing a loss function of energy and forces.
For an atomic configuration $\mathcal{C}$, the loss is
\begin{equation}
  l(\theta; \mathcal{C}) = w_\text{E} \left( \frac{E - \hat E}{N} \right)^2 + w_\text{F} \frac{\sum_{i=1}^N \| \bm F_i - \hat{\bm F_i} \|^2}{3N},
\end{equation}
where $N$ is the number of atoms, $E$  and $\hat E$ are the predicted and reference energies, respectively, and $\bm F_i$ and $\hat{\bm F_i}$ are the predicted and reference forces on atom $i$, respectively.
Both the energy weight $w_\text{E}$ and force weight $w_\text{F}$ are set to 1.
The total loss to minimize at each optimization step is the sum of the losses of multiple configurations,
\begin{equation}
   L(\theta)  = \sum_{j=1}^B l(\theta; \mathcal{C}_j),
\end{equation}
where $B$ is the mini-batch size.

The \model model is implemented in PyTorch~\cite{paszke2019pytorch} and trained with PyTorch Lightning~\cite{lightning}.
We trained all models using the Adam optimizer~\cite{kingma2014adam} with an initial learning rate of 0.01, which is reduced by a factor of 0.8 if the validation error does not decrease for 100 epochs.
The training batch size and allowed maximum number of training epochs vary for different datasets.
Training is stopped if the validation error does not reduce for a certain number of epochs (see Table~S4 in the \sishort).
We use an exponential moving average with a weight 0.999 to evaluate the validation set as well as for the final model.

Regarding model structure, Chebyshev polynomials of degrees up to $N_\beta = 8$ are used for the radial basis functions.
Other hyperparameters include the number of channels $u$, maximum tensor rank $v_\text{max}$, number of layers $T$, and cutoff distance $r_\text{cut}$.
Optimal values of these hyperparameters are searched for each dataset, and typical values are around $u=32$, $v_\text{max}=3$, $T=3$, and $r_\text{cut}=\SI{5}{\angstrom}$.
These result in small, parameter-efficient models with fewer than 125k parameters.
Detailed hyperparameters for each dataset are provided in Table~S5 in the \sishort.

\textbf{Diffusivity}.
The diffusivity can be computed from MD simulations via the Einstein equation, which  relates the MSD of particles to their diffusion coefficient:
\begin{equation}
  D = \lim_{t\rightarrow\infty} \frac{\left \langle  \frac{1}{N} \sum_i^N \left |\bm r_i(t) - \bm r_i(0) \right | ^{2} \right \rangle} {2nt},
\end{equation}
where the MSD is computed as the ensemble average ( denoted by $\langle \cdot \rangle$)
using diffusing atoms,
$\bm r_i(t)$ represents the position of atom $i$ at time $t$,
$N$ is the total number of considered diffusion atoms,
$n$ denotes the number of dimensions (three here),
and $D$ is the diffusion coefficient.
To solve for the diffusion coefficient $D$, we employ a linear fitting approach as implemented in ASE \cite{larsen2017atomic}, where $D$ is obtained as the slope of the MSD versus $2nt$.

The diffusivity of \ce{Li+} in LiPS was computed from MD simulations under the canonical ensemble using the Nos\'e--Hoover thermostat.
Simulations were performed at a temperature of \SI{520}{\kelvin} for a total of \SI{50}{\ps}, with a timestep of \SI{0.25}{\fs}, consistent with the settings reported in the benchmark study in \olcite{fu2023forces}.
We similarly computed the diffusivity of water; five simulations were performed at \SI{300}{\kelvin}, each using a timestep of \SI{1}{\fs} and running for a total of \SI{50}{\ps}.

\textbf{Stability criteria}.
For periodic systems, stability is monitored from the RDF $g(r)$ such that a simulation becoming ``unstable'' at time $t$ when \cite{fu2023forces}
\begin{equation}
  \int_{0}^\infty \|  \langle g(r)\rangle _{t}^{t+\tau} - \langle g(r) \rangle \| \, \text{d}r > \Delta,
\end{equation}
where $\langle\cdot\rangle$ denotes the average using the entire trajectory, $\langle \cdot \rangle _{t}^{t+\tau}$ denotes the average in a time window $[t, t+\tau]$, and $\Delta$ is a threshold.
In other words, when the difference of the area under the RDF obtained from the time window and the entire trajectory exceeds $\Delta$, the simulation is considered unstable.

This criterion cannot be applied to characterize the stability of MD simulations where large structural changes are expected, such as in phase transitions or chemical reactions.
However, for the LiPS system studied here, no such events are expected, and the RDF-based stability criterion is appropriate.
We adopted $\tau = \SI{1}{\ps}$ and $\Delta = 1$, as proposed in \olcite{fu2023forces}.

For molecular systems, the stability is monitored through the bond lengths, and a simulation is considered ``unstable'' at time $T$ when \cite{fu2023forces}
\begin{equation}
  \max_{(i,j)\in \mathcal{B}} | r^{ij}(T) - b^{ij} | > \Delta,
\end{equation}
where $\mathcal{B}$ denotes the set of all bonds, $\Delta$ is a threshold,
$r^{ij}(T)$ is the bond length between atoms $i$ and $j$ at time $T$, and $b^{ij}$ is the corresponding equilibrium bond length, computed as the average bond length using the reference DFT data.
For the MD17 dataset, we adopted $\Delta = 0.5$~\AA\ as in \olcite{fu2023forces}, and the MD simulations were performed at \SI{500}{\kelvin} for \SI{300}{\ps} with a timestep of \SI{0.5}{\fs}, using the Nos\'e--Hoover thermostat.

\section*{Data availability}

The new bilayer graphene dataset is provided at \url{https://github.com/wengroup/camp_run}.
The other datasets used in this work are publicly available;
the LiPS dataset: \url{https://archive.materialscloud.org/record/2022.45},
the water dataset: \url{https://doi.org/10.1073/pnas.1815117116},
and the MD17 dataset: \url{http://www.sgdml.org}.

\section*{Code availability}
The code for the \model model is available at \url{https://github.com/wengroup/camp}.
Scripts for training models, running MD simulations, and analyzing the results are at \url{https://github.com/wengroup/camp_run}.

\section*{Author contributions}
M.W.: project conceptualization, model development, data analysis, writing - original draft, writing - review, and supervision.
W.F.H, J.D. and S.A.: data analysis and writing - review.

\section*{Conflicts of interest}
There are no conflicts of interest to declare.

\section*{Acknowledgements}
This work is supported by the National Science Foundation under Grant No.\ 2316667, and supported by the Center for HPC at the University of Electronic Science and Technology of China. 
This work also uses computational resources provided by the Research Computing Data Core at the University of Houston.

\def\bibsection{\section*{\refname}} 
%

\end{document}


\title{\Large Supplementary Information:\\Cartesian Atomic Moment Machine Learning Interatomic Potentials}

\author{Mingjian Wen$^1$}
\email{mjwen@uestc.edu.cn}
\author{Wei-Fan Huang$^2$}
\author{Jin Dai$^2$}
\author{Santosh Adhikari$^2$}
\affiliation{$^1$Institute of Fundamental and Frontier Sciences, University of Electronic Science and Technology of China, Chengdu, 611731, China \\
$^2$Department of Chemical and Biomolecular Engineering, University of Houston, Houston, TX, 77204, USA
}

\maketitle

\section{Atomic Moment tensor construction example}
\label{sec:atomic:moment}

Here, we give concrete examples of constructing the atomic moment tensors $\bm M$ up to a rank of $v_\text{max}=2$.
Higher-rank ones can be constructed similarly.

Let's start by considering the atomic moment in Eq.~(5) in the main text.
For ease of discussion, we copy the equation below:
\begin{equation} \label{eq:atomic:moment}
    \bm M^i_{uv, p} = \sum_{j \in \mathcal{N}_i} R_{uvv_1v_2} \bm h^j_{uv_1} \contract \bm D^{ij}_{v_2}.
\end{equation}
The tensor ranks are only determined by the subscript $v_1$, $v_2$, and $v$, and are not affected by either the radial function $R$ or the summation over neighbors.
Therefore, to simplify the presentation, we consider the below equation that captures the essence of ~\eref{eq:atomic:moment}:
\begin{equation} \label{eq:atomic:moment:simple}
    \bm M_{v} = \bm h_{v_1} \contract \bm D_{v_2},
\end{equation}
where $v_1$  denotes the rank of $\bm h$ and $v_2$ denotes the rank of $\bm D$, $v$ denotes the rank of $\bm M$, and $\contract$ denotes the tensor contraction between the two tensors, with $c$ being the contraction degree.
If we choose to set the maximum allowed tensor rank to $v_\text{max}=2$, then we can have features up to rank 2 ($\bm h_0$, $\bm h_1$, and $\bm h_2$) and generalized dyadic tensors up to rank 2 as well ($\bm D_0$, $\bm D_1$, and $\bm D_2$).

The atomic moment tensor $\bm M$ of different ranks are constructed by following two rules:
\begin{enumerate}
    \item the rank of the feature tensor $\bm h_{v_1}$ should be smaller than or equal to the rank of the dyadic tensor $\bm D_{v_2}$, i.e., $v_1 \leq v_2$; and
    \item the indices of the feature tensor $\bm h_{v_1}$ needs to be completely contracted away, i.e., $ c=v_1$.
\end{enumerate}
These design principles ensure that information passes naturally from higher-rank tensors to lower-rank tensors, which is essential for modeling the potential energy surface, a scalar quantity.
It also reduces the number of high-rank tensors, making the model more computationally efficient.

$\bm M_0$ tensors (scalars) are constructed from the below tensors:
\begin{itemize}[label={}]
    \setlength{\itemsep}{0pt}
    \item $\bm h_0 \contractc{0} \bm D_0$  \quad (product of two scalars)
    \item $\bm h_1 \contractc{1} \bm D_1$  \quad (dot product of two vectors)
    \item $\bm h_2 \contractc{2} \bm D_2$  \quad (2-contraction between two second rank tensors, i.e. $\sum_{ij} h^{ij}D^{ij}$, where $i$ and $j$ are the tensor indices)
\end{itemize}
All three operations produce scalars.
As mentioned in the main text, each of them is called a path and they are linearly combined to form the final atomic moment tensor $\bm M_0$.

Similarly, $\bm M_1$ tensors (vectors) are constructed from the below tensors:
\begin{itemize}[label={}]
    \setlength{\itemsep}{0pt}
    \item $\bm h_0 \contractc{0} \bm D_1$ \quad (scalar multiplied by a vector)
    \item $\bm h_1 \contractc{1} \bm D_2$ \quad (1 contraction between a vector and a second rank tensor, i.e., $\sum_{i} h^{i}D^{ij}$)
\end{itemize}
The two operations produce vectors and are linearly combined to form the final atomic moment tensor $\bm M_1$.

Finally, $\bm M_2$ tensors (second rank tensors) are constructed from the below tensors:
\begin{itemize}[label={}]
    \setlength{\itemsep}{0pt}
    \item $\bm h_0 \contractc{0} \bm D_2$ \quad (scalar multiplied by a second rank tensor)
\end{itemize}
It produces a second-rank tensor, which is the final atomic moment tensor $\bm M_2$.

\textbf{We emphasize that, in all the above contractions, it does not matter which indices are used for contraction because both $\bm h$ and $\bm D$ are symmetric tensors.
For example, $\sum_{i} h^{i}D^{ij} = \sum_{j} h^{j}D^{ij}$.
}

\section{Hyper moment tensor construction example}
\label{sec:hyper:moment}

Here, we provide an example of constructing the hyper moment tensor $\bm H$ up to a rank of $v_\text{max}=3$.
We start with Eq.~(7) in the main text (copied below for convenience):
\begin{equation} \label{eq:hyper:moment}
    \bm H_{uv,p} = \bm M_{uv_1} \contract \bm M_{uv_2} \contract ... \contract \bm M_{uv_n}.
\end{equation}
Similar to the atomic moment tensor construction, we consider the simplified version of \eref{eq:hyper:moment}:
\begin{equation} \label{eq:hyper:moment:simple}
    \bm H_v = \bm M_{v_1} \contractc{c_1} \bm M_{v_2} \contractc{c_2} \dots \contractc{c_{n-1}} \bm M_{v_n}
\end{equation}
where $v_1$, $v_2$, ..., $v_n$ denote the ranks of the atomic moment tensors $\bm M$ and $v$ denotes the rank of output the hyper moment.
For a maximum allowed tensor rank of $v_\text{max}=3$, we can have atomic moment tensors up to rank 3 ($\bm M_0$, $\bm M_1$,  $\bm M_2$, and $\bm M_3$).

Similar to the atomic moment tensor construction, the hyper moment tensors $\bm H$ of different ranks are constructed by following two rules:
\begin{enumerate}
    \item All contractions are between a previous tensor $M_{v_1}, \dots, M_{v_{n-1}}$ and the last tensor $M_{v_n}$ in the sequence.
          This implies that the rank of the previous tensor should be smaller than or equal to the rank of the last tensor, i.e., $v_i \leq v_n$, where $i=1,2,\dots,n-1$.
    \item The indices of the previous tensor $M_{v_1}, \dots, M_{v_{n-1}}$ need to be completely contracted away, i.e., $c_i=v_i$.
\end{enumerate}
Again, as mentioned in the main text, the guiding principle for constructing the hyper moment tensor is to result in the information flow from higher-rank tensors to lower-rank tensors.
This is driven by the fact that we are modeling the potential energy surface, which is a scalar quantity.
An added benefit is that this can also reduce the number of high-rank tensors, which can be computationally expensive to compute and store.

The hyper moment tensor $\bm H$ of different ranks are constructed as follows.

$\bm H_0$ tensors (scalars) are constructed from the below tensors:
\begin{itemize}[label={}]
    \setlength{\itemsep}{0pt}
    \item $\bm M_0$  \quad (body order 2)
    \item $\bm M_1 \contractc{1} \bm M_1$ \quad (body order 3)
    \item $\bm M_2 \contractc{2} \bm M_2$ \quad (body order 3)
    \item $\bm M_3 \contractc{3} \bm M_3$ \quad (body order 3)
    \item $\bm M_1 \contractc{1} \bm M_1 \contractc{1} \bm M_2$ \quad (body order 4)
    \item $\bm M_1 \contractc{1} \bm M_2 \contractc{2} \bm M_3$ \quad (body order 4)
    \item $\bm M_1 \contractc{1} \bm M_1 \contractc{1} \bm M_1 \contractc{1} \bm M_3$ \quad (body order 5)
\end{itemize}
All seven operations produce scalars and are linearly combined to form the final hyper moment tensor $\bm H_0$.
As mentioned in the main text, it is required that all contractions are between
a previous tensor and the last tensor in the sequence.
For example, in $\bm M_1 \contractc{1} \bm M_2 \contractc{2} \bm M_3$, the double contraction between $\bm M_2$ and $\bm M_3$ is carried out first, and then the resultant vector is single-contracted with $\bm M_1$, i.e., $\sum_{ijk} M^{i}
    M^{jk} M^{ijk}$, where $i$, $j$, and $k$ are the tensor indices.
Again, the order of the indices being selected for contractions does not matter because all the tensors are symmetric, for example,
$\sum_{ijk} M^{i} M^{jk} M^{ijk} = \sum_{ijk} M^{k} M^{ij} M^{ijk}$.

Any number of atomic moment tensors can be used to construct a hyper moment tensor.
This choice enables quick body order expansion of the potential energy surface.
For example, in the first layer, each atomic moment tensor $\bm M$, irrespective of its rank, only captures two-body interactions.
This is because the atomic moment is constructed from the scalar features and the dyadic tensors $\bm D$, which, in turn, depends only on an atom $i$ and its neighbor $j$, involving only two atoms.
With the moment tensor $\bm H_0$ constructed above from the atomic moment tensors $\bm M$, the interaction body orders can reach 5 with only a single layer.
This is analogous to the B basis used in ACE and MACE.

$\bm H_1$ tensors (vectors) are constructed from the below tensors:
\begin{itemize}[label={}]
    \setlength{\itemsep}{0pt}
    \item $\bm M_1 $
    \item $\bm M_1 \contractc{1} \bm M_2$
    \item $\bm M_2 \contractc{2} \bm M_3$
    \item $\bm M_1 \contractc{1} \bm M_1 \contractc{1} \bm M_3$
\end{itemize}

$\bm H_2$ tensors (second rank tensors) are constructed from the below tensors:
\begin{itemize}[label={}]
    \setlength{\itemsep}{0pt}
    \item $\bm M_2$
    \item $\bm M_1 \contractc{1} \bm M_3$
\end{itemize}

$\bm H_3$ tensors (third rank tensors) are constructed from the below tensors:
\begin{itemize}[label={}]
    \setlength{\itemsep}{0pt}
    \item $\bm M_3$
\end{itemize}

\section{Complexity analysis}
The time complexity of the \model model is dominated by the construction of the atomic moment tensors and the hyper moment tensors.
As detailed in the above two sections, the atomic moment tensors are created by contracting two tensors as in \eref{eq:atomic:moment:simple}.
For this contraction, we have $v = v_1 + v_2 - 2c$, where $c$ is the contraction degree.
The time complexity is $O(3^{v_1+v_2-c})$.
In \sref{sec:atomic:moment}, we have discussed that \model has the constraint that $c=v_1$.
Then the time complexity of the atomic moment tensor is $O(3^{v_2})$.
Let the maximum rank of the tensors in \model be $v_\text{max}$, then tensors up to rank $v_\text{max}$ can be used in the atomic moment tensor construction.
So, $v_2 \leq v_\text{max}$.
\textbf{Therefore, the time complexity of the atomic moment tensor construction is $O(3^{v_\text{max}})$.}
For example, when $v_\text{max}=2$ as discussed in \sref{eq:atomic:moment}, the operations of $O(3^2)$ complexity are those involving $\bm D_2$.

The hyper moment tensors are constructed by contracting multiple tensors as in
\eref{eq:hyper:moment:simple}.
In \sref{sec:hyper:moment}, we have discussed that \model has the constraint that $c_i=v_i$, and every contraction is between a previous tensor and the last tensor in the sequence.
Therefore, in terms of the time complexity analysis, \eref{eq:hyper:moment:simple} is equivalent to
\begin{equation}
    \bm H_v = \bm T_{v'} \contractc{c} \bm M_{v_n},
\end{equation}
where $\bm T_{v'}$ represents the sequence of tensors $\bm M_{v_1}, \dots, \bm M_{v_{n-1}}$, $v'= v_1 + \dots + v_{n-1}$, and $c = v'$.
The time complexity of this contraction is $O(3^{v'+v_n - c}) = O(3^{v_n})$.
Given that the maximum rank of the tensors in \model is $v_\text{max}$, we have $v_n \leq v_\text{max}$.
\textbf{Therefore, the time complexity of the hyper moment tensor construction is $O(3^{v_\text{max}})$.}
For example, when $v_\text{max} = 3$ as discussed in \sref{sec:hyper:moment}, the operations of $O(3^3)$ complexity are those involving $\bm M_3$ as the last tensor in \eref{eq:hyper:moment:simple}.

\section{Extra results}

\begin{figure}[H]
    \centering
    \includegraphics[width=0.4\textwidth]{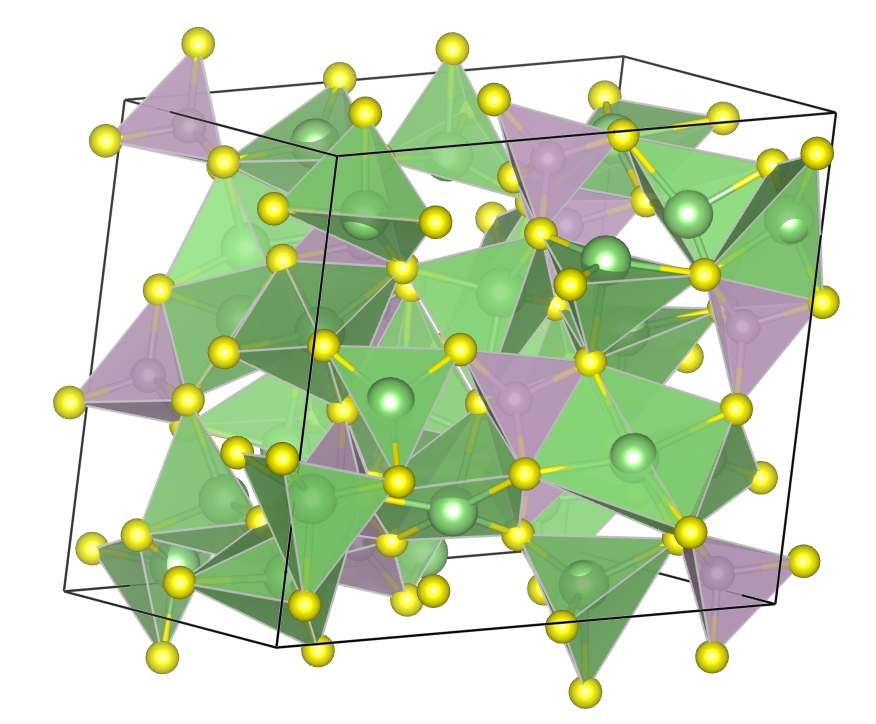}
    \caption{Structure of the LiPS (\ce{Li_{6.75}P3S11}) crystal. The Li, P, and S atoms are shown in green, purple, and yellow, respectively.
        For the angular distribution function, the S-P-S angles in the purple tetrahedra are used.
    }
    \label{fig:lips:struct}
\end{figure}

\begin{table}[h]
    \centering

    \footnotesize
    \caption{Running speed of various models (DeepPot-SE \cite{deeppotse}, SchNet \cite{schutt2018schnet}, DimeNet \cite{gasteiger2020dimenet}, PaiNN \cite{painn}, SphereNet \cite{liu2022spherical}, ForceNet \cite{hu2021forcenet}, GemNet-T \cite{gemnet}, GemNet-dT, NequIP \cite{batzner2022e}) for a water system with 192 atoms.
        The speed is measured as the number of frames per second (FPS) in an MD simulation using a single NVIDIA V100 GPU.
        All results, except for CACE \cite{cheng2024cartesian} and \model, are obtained from water-1k results in \olcite{fu2023forces}.
        Although these models are trained on a water dataset different from that used for CACE and \model, model running speed for water with 192 atoms are comparable.
    }
    \resizebox{\columnwidth}{!}{
    \begin{tabular}{cccccccccccccc}
        \hline
         DeepPot-SE  & SchNet  & DimeNet  & PaiNN  & SphereNet  & ForceNet  & GemNet-T  & GemNet-dT & NequIP  & CACE  & \model \\

        \hline
        61.8       & 99.2   & 16.4    & 54.5  & 3.0       & 68.1     & 15.4     & 34.5      & 3.9  &9.8 &14.2                   \\
        \hline
    \end{tabular}}
    \label{table:fps_water1k}
\end{table}

\begin{figure}
    \centering
    \includegraphics[width=0.5\textwidth]{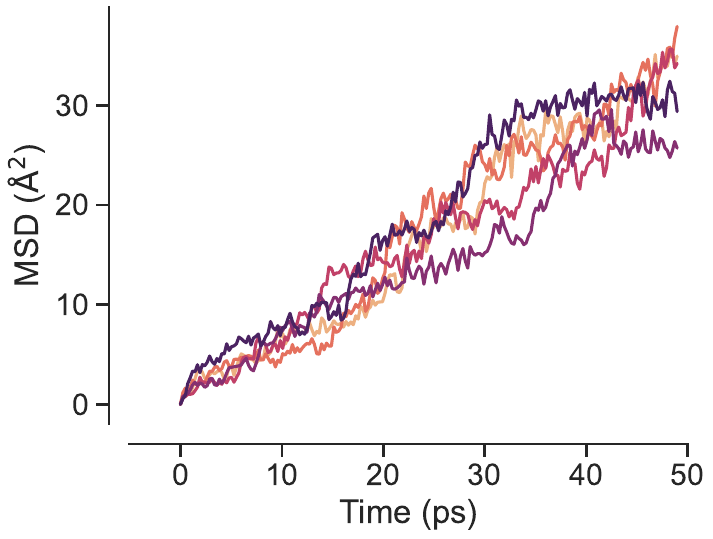}
    \caption{MSD of LiPS using \model at a large timestep of 1~fs.
        The computed diffusivity is $D = (1.11\pm0.11) \times10^{-5}$~\si{\cm^2\per\s}.
    }
    \label{fig:LiPS:msd:tau1}
\end{figure}

\begin{figure}
    \centering
    \begin{subfigure}[b]{0.5\textwidth}
        \includegraphics[width=1\textwidth]{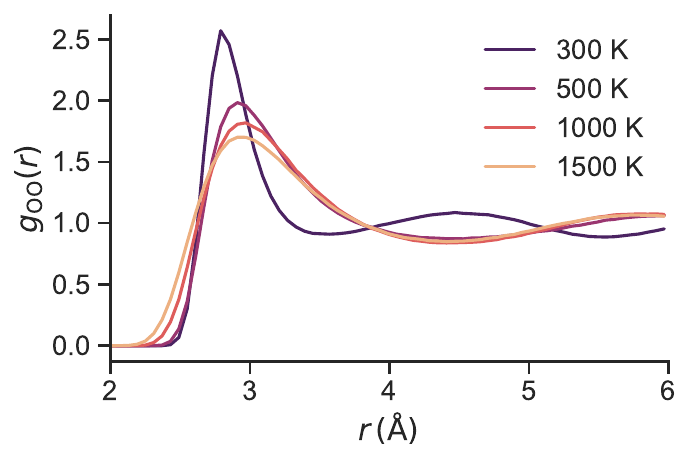}
        \caption{}
        \label{fig:water:rdf:highT}
    \end{subfigure}\\
    \begin{subfigure}[b]{0.8\textwidth}
        \includegraphics[width=1\textwidth]{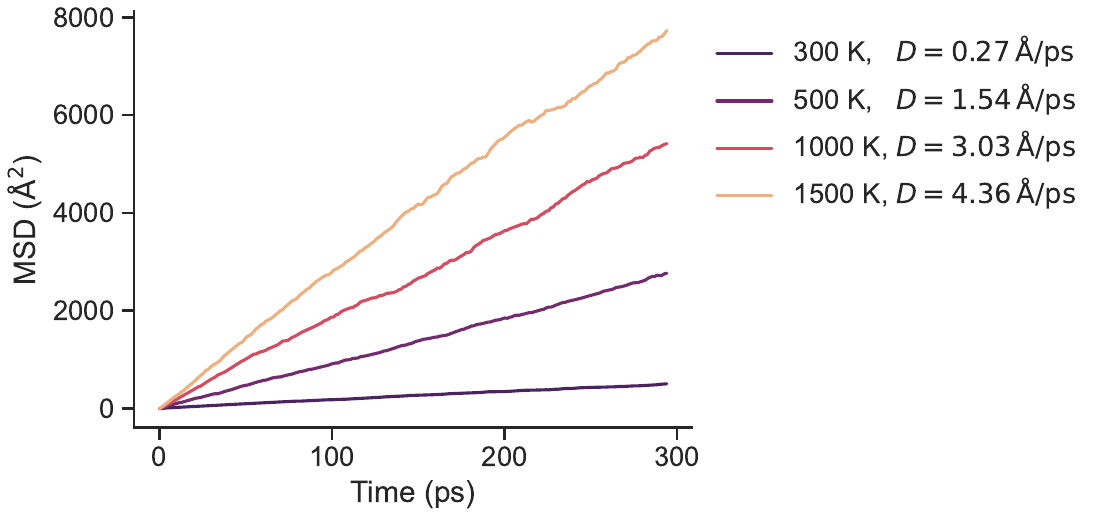}
        \caption{}
        \label{fig:water:msd}
    \end{subfigure}
    \caption{
        MD simulation results of 512 water molecules using \model at various temperatures.
        (a) Radial distribution function.
        (b) Mean square displacement and the corresponding diffusion coefficient.
    }
\end{figure}

\begin{table}[H]
    \footnotesize
    \caption{
        Results of Aspirin and Ethanol trained on 9500 structures.
        Energy MAE in meV, force MAE in meV/\AA, stability measured in picoseconds (larger the better).
        Results from all models (DeepPot-SE \cite{deeppotse}, SchNet \cite{schutt2018schnet}, DimeNet \cite{gasteiger2020dimenet}, PaiNN \cite{painn}, SphereNet \cite{liu2022spherical}, ForceNet \cite{hu2021forcenet}, GemNet-T \cite{gemnet}, and NequIP \cite{batzner2022e}) except for \model, are obtained from benchmark study in \olcite{fu2023forces}.
    }
    \begin{tabular}{llcccccccccccc}
        \hline
        Molecule & Metric    & DeepPot-SE  & SchNet        & DimeNet     & PaiNN         & SphereNet    & ForceNet      & GemNet-T     & NequIP      & \model      \\
        \hline
        Aspirin  & Energy    &             &               &             &               &              &               &              &             & 5.0         \\
                 & Forces    & 21.0        & 35.6          & 10.0        & 9.2           & 3.4          & 22.1          & 3.3          & 2.3         & 5.5         \\
                 & Stability & 9$_{(15)}$  & 26$_{(23)}$   & 54$_{(12)}$ & 159$_{(121)}$ & 141$_{(54)}$ & 182$_{(144)}$ & 72$_{(50)}$  & 300$_{(0)}$ & 300$_{(0)}$ \\
                 & Speed     & 88.0        & 108.9         & 20.6        & 85.8          & 17.5         & 137.3         & 28.2         & 8.4         & 35.0        \\
        \hline
        Ethanol  & Energy    &             &               &             &               &              &               &              &             & 2.1         \\
                 & Force     & 8.9         & 16.8          & 4.2         & 5.0           & 1.7          & 14.9          & 2.1          & 1.3         & 2.1         \\
                 & Stability & 300$_{(0)}$ & 247$_{(106)}$ & 26$_{(10)}$ & 86$_{(109)}$  & 33$_{(16)}$  & 300$_{(0)}$   & 169$_{(98)}$ & 300$_{(0)}$ & 300$_{(0)}$ \\
                 & Speed     & 101.0       & 112.6         & 21.4        & 87.3          & 30.5         & 141.1         & 27.1         & 8.9         & 32.8        \\
        \hline
    \end{tabular}
\end{table}

\begin{table}[H]
\centering
\caption{MAEs on aspirin and ethanol from the rMD17 dataset \cite{christensen2020Oct}.
Energy (meV) and force (meV/\AA) errors are reported.
Models are trained on 950 configurations, validated on 50 configurations, and tested on 1000 configurations.
All results are from \olcite{batatia2022mace}, except for CAMP.
}
\label{tab:results}
\begin{tabular}{cccccccccc}
\hline
Molecule & Metric & GAP & FCHL & ACE & PaiNN & NequIP & Allegro & MACE & CAMP \\
\hline
Aspirin & Energy & 17.7 & 6.2 & 6.1 & 6.9 & 2.3 & 2.3 & 2.2 & 5.6 \\
        & Forces & 44.9 & 20.9 & 17.9 & 16.1 & 8.2 & 7.3 & 6.6 & 17.1\\
\hline
Ethanol  & Energy & 3.5 & 0.9 & 1.2 & 2.7 & 0.4 & 0.4 & 0.4 & 0.7 \\
        & Forces & 18.1 & 6.2 & 7.3 & 10.0 & 2.8 & 2.1 & 2.1 & 5.0 \\
\hline
\end{tabular}
\end{table}

\section{Training details}

\setlength{\tabcolsep}{6pt} 
\begin{table}[H]
    \centering
    \caption{Training hyperparameters for the \model models.}
    \begin{tabular}{cccc}
        \hline
        Dataset
                         & Allowed epochs
                         & Early stopping patience
                         & Batch size                        \\
        \hline
        LiPS             & 5000                    & 400 & 2 \\
        Water            & 10000                   & 400 & 4 \\
        MD17             & 10000                   & 800 & 4 \\
        Bilayer graphene & 1000                    & 200 & 4 \\
        \hline
    \end{tabular}
\end{table}

\begin{table}[H]
    \caption{Optimal architectural hyperparameters for the \model models.
        The optimal hyperparameters are determined by a grid search:
        $u\in [16, 32, 48, 64]$,
        $v_\text{max}\in [2,3]$,
        $T\in [1,2,3]$, and
        $r_\text{cut}\in [5, 5.5, 6]$.
    }
    \label{tab:opt:arch}
    \begin{tabular}{cccccc}
        \hline
        Dataset          & Training set size
                         & \# channels
                         & Max tensor rank
                         & \# layers
                         & Cutoff                                \\
                         &
                         & $u$
                         & $v_\text{max}$
                         & $T$
                         & $r_\text{cut}$ (\AA)                  \\
        \hline
        LiPS             & 10                   & 16 & 3 & 2 & 6 \\
        LiPS             & 100                  & 16 & 3 & 2 & 6 \\
        LiPS             & 1000                 & 48 & 3 & 3 & 6 \\
        LiPS             & 2500                 & 64 & 3 & 3 & 6 \\
        LiPS             & 19000                & 48 & 3 & 3 & 6 \\
        Water            & 1433                 & 32 & 2 & 3 & 5 \\
        Aspirin          & 950                  & 48 & 3 & 2 & 6 \\
        Ethanol          & 950                  & 48 & 3 & 2 & 6 \\
        Malonaldehyde    & 950                  & 32 & 3 & 3 & 6 \\
        Naphthalene      & 950                  & 32 & 3 & 3 & 6 \\
        Salicylic aci    & 950                  & 48 & 3 & 2 & 5 \\
        Toluene          & 950                  & 48 & 3 & 2 & 6 \\
        Uracil           & 950                  & 48 & 3 & 3 & 5 \\
        Bilayer graphene & 5560                 & 48 & 3 & 3 & 6 \\
        \hline
    \end{tabular}
\end{table}

\section{Ablation study}

We conducted an ablation study to evaluate the impact of \model's architectural design on prediction accuracy.
The study was performed using the LiPS 100 dataset, and all reported results correspond to the test set.

First, we examined the effect of the rank of the internal Cartesian tensor representation.
We found that the model utilizing tensorial features outperforms the invariant model that relies solely on scalar features.
When \verb|max_v| is set to 0, \model reduces to an invariant model using only scalar features, achieving an energy error of 2.12~meV.
(This represents the lowest error achievable when fixing \verb|max_v| to 0 while optimizing other hyperparameters. The same applies to the results below.)
By transitioning to an equivalent model with vector features (i.e., \verb|max_v=1|), the energy error decreases to 1.17~meV.
Further improvements are observed with higher-rank tensorial features, yielding energy errors of 0.62, 0.43, and 0.43~meV for \verb|max_v=2, 3|, and \verb|4|, respectively.
These results indicate that tensorial features are crucial for achieving high model performance.
Empirically, features of rank 3 appear to be a suitable choice, as demonstrated throughout this work and summarized in \tref{tab:opt:arch}.

Second, we found that message passing is essential for enhancing model performance.
Without message passing, \model with a single layer behaves similarly to the MTP model \cite{shapeev2016moment}, achieving an energy error of 0.84~meV.
However, when message passing is enabled, the energy error improves significantly, reaching 0.43~meV and 0.45~meV for two and three layers, respectively.

Finally, we observed that other parameters have a relatively minor impact compared to the two factors discussed above, consistent with existing findings.
A cutoff distance in the range of $5\sim6$~\AA\ results in energy errors around 0.43~meV, whereas a smaller value, such as 4~\AA, leads to a significantly higher energy error of 1.06~meV.
The number of radial channels (\verb|max_u|) shows diminishing returns once it exceeds 16, with an energy error of 0.43~meV.
Similarly, the model's performance saturates when the embedding dimension for chemical species (\verb|max_chebyshev_degree|) exceeds 8.

\def\bibsection{\section*{\refname}} 
%